\documentclass[showpacs,preprintnumbers,preprint,amsmath,amssymb]{revtex4}

\usepackage{graphicx}
\usepackage{dcolumn}
\usepackage{bm}
\begin{document}

\preprint{LA-UR 05-7372}

\title{A Model for Transits in Dynamic Response Theory}

\author{Giulia De Lorenzi-Venneri}
\author{Duane C. Wallace}
\affiliation{Theoretical Division, Los Alamos National Laboratory, 
Los Alamos, New Mexico 87545}

\date{\today}

\begin{abstract}
The first goal of Vibration-Transit (V-T) theory was to construct a tractable approximate
Hamiltonian from which the equilibrium thermodynamic properties
of monatomic liquids can be calculated. The Hamiltonian for vibrations  in an
infinitely extended harmonic random valley, together with the universal
multiplicity of such valleys, gives an accurate first-principles account of the
measured thermodynamic properties of the elemental liquids at melt. In the
present paper, V-T theory is extended to non-equilibrium properties, through
an application to the dynamic structure factor $S(q,\omega)$. It was previously
shown that the vibrational contribution alone accurately accounts for the
Brillouin peak dispersion curve for liquid sodium, as compared both with
MD calculations and inelastic x-ray scattering data. Here it is argued that
the major effects of transits will be to disrupt correlations within the
normal mode vibrational motion, and to provide an additional source of
inelastic scattering. We construct a parameterized model for these effects,
and show that it is capable of fitting MD results for $S(q,\omega)$ in liquid
sodium. A small discrepancy between model and MD at large $q$ is attributed to
multimode vibrational scattering. In comparison, mode coupling theory
formulates $S(q,\omega)$ in terms of processes through which
density fluctuations decay. While mode coupling theory is also capable of
modeling $S(q,\omega)$ very well, V-T theory is the more universal since it
expresses all statistical averages, thermodynamic functions and time 
correlation functions alike, in terms of the same motional constituents,
vibrations and transits.

\end{abstract}

\pacs{05.20.Jj, 63.50.+x, 61.20.Lc, 61.12.Bt}
\keywords {Liquid Dynamics, Inelastic Neutron Scattering, Dispersion Relations, Mode Coupling Theory,
V-T Theory}
\maketitle

\section{Introduction}

Vibration-Transit (V-T) theory is a Hamiltonian formulation of the dynamics 
of monatomic liquids. It is based on the idea that a liquid system moves on
a potential energy surface making jumps between valleys, that these jumps are approximately instantaneous,
and that the dominant majority of visited valleys are all random in structure
and are equivalent in energy and vibrational properties. The zeroth order approximation to the 
Hamiltonian expresses the liquid motion in terms of normal mode
vibrations  in a single infinitely-extended harmonic random 
valley, and can be explicitly calculated from first principles for actual systems. 
It is now well known that this vibrational motion gives a very good account of the 
equilibrium thermodynamics of monatomic elemental liquids at melt \cite{DWPRE56,ChWJPCM01,DWbook2}. 
This result is conceptually fundamental,
not only because it supports the potential energy landscape picture of liquid dynamics,
but also because it validates the basic assumptions of V-T theory. Moreover, it was obtained by V-T theory
without adjustable parameters, a result that no other tractable theory has achieved. 
Following this success, in the present work we look into a deeper level of the dynamical behavior 
in liquids, namely its nonequilibrium properties,  and apply V-T theory 
to time correlation functions, which express nonequilibrium properties in
linear response theory \cite{HMCDbook}. Again without adjustable parameters, the
vibrational contribution to any time correlation function can be calculated from
the zeroth order Hamiltonian. Once more the vibrational contribution turns out to 
play a central role and precisely gives
 the location of the Brillouin peak in the inelastic 
scattering data for liquid sodium \cite{ARXIV05}. However, the width of the 
Brillouin peak is larger than the vibrational width alone. This
broadening of the Brillouin peak results from the transit contribution, for which  an explicit
evaluation is not yet available. The purpose of this paper
 is to construct and test a model for
the contribution of transits to inelastic scattering. The model is shown to be very 
successful for our case study,
and thus  provides a new description of the scattering process.

The dynamics of liquids and supercooled liquids, studied with the aid of
 MD calculations for glass forming systems, is a currently active research
 field. We have presented extensive comparison of V-T theory with a broad range
 of potential-energy-landscape theories \cite{ChDW05}. In a paper of particular relevance here,
 Mazzacurati, Ruocco, and Sampoli
 \cite{Mazz96} (see also \cite{R&cPRL00}) have shown that a vibrational analysis is in
 excellent agreement with MD calculations of $S(q,\omega)$ for a Lennard-Jones 
 glass. Beyond this result, extensive theoretical analysis
 of the complete atomic motion is required before a vibrational contribution
 can be incorporated into a theory of liquid dynamics. This analysis 
 contitues the foundation of V-T theory \cite{DWPRE56,ChWJPCM01,DWbook2} and provides
 the following stipulations: (a) potential energy valleys used in liquid theory 
 must be random valleys, and not some other symmetry; (b) because all random
 valleys of a given system are equivalent in vibrational properties, the liquid
 vibrational contribution can be calculated from a single random valley; 
 (c) the representative random
 valley has to be extended to infinity so that the vibrational statistical 
 averages are defined;  (d) for thermodynamic functions, corrections for anharmonicity 
 and valley-valley intersections must be
 recognized; and (e) for time correlation functions, 
 the vibrational motion 
 has to be supplemented with transits in the liquid.  These stipulations
 are crucial to the present theoretical development.

Starting from the exact first-principles vibrational contribution, our
model for transits in dynamic response theory is constructed in Sec.\;II.
In Sec.\;III, the model parameters are adjusted to achieve agreement with
MD calculations for a system representing liquid sodium at melt. The
quality of the fitting is discussed, as well as the interpretation of the
fitted parameters.  For this example
of dynamic response in monatomic liquids, we present in Sec.\;IV a detailed comparison 
of our V-T theory with mode coupling theory, which is at present the  most successful in 
accounting for time correlation functions in liquids. Our conclusions are
summarized in Sec.\;V, and the unifying nature of V-T theory is noted.

\section{Construction of the Transit Model}

Let us consider a system of $N$ atoms in a cubic box with the motion
governed by periodic boundary conditions. The position of atom $K$ at time
$t$ is $\bm{r}_{K}(t)$,  $K=1, \dots,N$. The density autocorrelation function is 
\begin{equation} \label{eq1}
F(q,t) =  \frac {1}{N} \left< \sum_{K} e^{-i \bm{q}\cdot \bm{r}_{K}(t)} 
                                       \sum_{L} e^{i \bm{q}\cdot \bm{r}_{L}(0)}
            \right >,
\end{equation}
where  $\left < \dots \right >$ represents a thermal average over the motion, plus an average over
the star of $\bm{q}$, which converts the right side to a function of $q$ for finite
systems. In V-T theory of the liquid state, the motion consists of normal
mode vibrations within a single extended (harmonic) random valley, 
plus transits between valleys. We shall neglect anharmonicity, and will
consider classical motion so that position coordinates may be commuted  at
will.

Let us neglect transits for the moment, and consider motion in a single
random valley. It is convenient to write
\begin{equation}\label{eq2}
\bm{r}_{K}(t) = \bm{R}_{K} + \bm{u}_{K}(t),
\end{equation}
where $\bm{R}_{K}$ is the equilibrium position and $\bm{u}_{K}(t)$ the displacement. The
contribution to $F(q,t)$ is the vibrational contribution, given by
\begin{equation} \label{eq3}
F_{vib}(q,t) = \frac{1}{N}
 \sum_{KL} e^{-i \bm{q}\cdot \bm{R}_{KL}} 
 \left < e^{-i \bm{q} \cdot (\bm{u}_{K}(t) - \bm{u}_{L}(0))} \right >,
\end{equation}
where $\bm{R}_{KL}=\bm{R}_{K}-\bm{R}_{L}$. The motional average is now a harmonic vibrational 
average $\left < \dots \right >_{h}$, and Eq.~(\ref{eq3}) simplifies to (see e.g. \cite{ChDW05})
\begin{equation}\label{eq4}
F_{vib}(q,t)=
\frac {1}{N} \Bigg < \sum_{KL}e^{-i \bm{q}\cdot \bm{R}_{KL}}
\; e^{-W_{K}(\bm{q})}e^{-W_{L}(\bm{q})}
\;\left[ 1 + \left < \bm{q} \cdot \bm{u}_{K}(t)\;\bm{q} \cdot \bm{u}_{L}(0)\right >_{h}+ \dots\right]
                     \Bigg >_{\bm{q}^{\ast}} ,
\end{equation}
where $W_{K}(\bm{q})$ is the Debye-Waller factor for atom $K$,
\begin{equation}\label{eq5}
W_{K}(\bm{q})=\frac{1}{2}\left < (\bm{q}\cdot\bm{u}_{K})^{2}\right  >_{h},
\end{equation}
and where  $\left < \dots \right >_{q^{\ast}}$ is the $\bm{q}$-star average. The series in brackets 
in Eq.~(\ref{eq4}) is the expansion of an exponential. Since 
$\left < \bm{q} \cdot \bm{u}_{K}(t)\;\bm{q} \cdot \bm{u}_{L}(0)\right >_{h}$ vanishes as
$t\rightarrow \infty$, the constant term in Eq.~(\ref{eq4}) is $F_{vib}(q,\infty)$, given by
\begin{equation} \label{eq6}
F_{vib}(q,\infty) = \frac {1}{N} \left < \sum_{KL}\cos(\bm{q}\cdot \bm{R}_{KL})\;e^{-W_{K}(\bm{q})} e^{-W_{L}(\bm{q})}
                     \right >_{\bm{q}^{\ast}},
\end{equation}
where $\cos (\bm{q}\cdot \bm{R}_{KL})$ appears because of the
star average. To leading order in the expansion in Eq.~(\ref{eq4}), the time dependence
of $F_{vib}(q,t)$ is contained in the function 
\begin{equation}\label{eq7}
F_{vib}(q,t)-F_{vib}(q,\infty)=
\frac {1}{N} \Bigg < \sum_{KL}\cos(\bm{q}\cdot \bm{R}_{KL})  
\;e^{-W_{K}(\bm{q})} e^{-W_{L}(\bm{q})}
\;\left < \bm{q} \cdot \bm{u}_{K}(t)\;\bm{q} \cdot \bm{u}_{L}(0)\right >_{h}
                     \Bigg >_{\bm{q}^{\ast}} .
\end{equation}

To evaluate the dynamic structure factor $S(q,\omega)$, the displacements
are written as a sum over normal modes $\lambda$, which have frequencies $\omega_{\lambda}$ and
eigenvector components $\bm{w}_{K\lambda}$, for $\lambda=1, \dots,3N$. The result is \cite{ChDW05}
\begin{equation}\label{eq8}
S_{vib}(q,\omega) = F_{vib}(q,\infty)\delta(\omega)+S_{vib}^{(1)}(q,\omega),
\end{equation}
where 
\begin{equation} \label{eq9}
S_{vib}^{(1)}(q,\omega)=\frac{3kT}{2M}\frac{1}{3N}\sum_{\lambda}f_{\lambda}(q)[\delta(\omega+\omega_{\lambda})+
\delta(\omega-\omega_{\lambda})], 
\end{equation}
\begin{equation} \label{eq10}
f_{\lambda}(q)=\frac{1}{\omega_{\lambda}^{2}}\Bigg < \left|\sum_{K}e^{-i \bm{q}\cdot {\bm{R}}_{K} } 
\;e^{-W_{K}(\bm{q})} 
 \bm{q} \cdot \bm{w}_{K\lambda}
                     \right|^{2}\Bigg >_{\bm{q}^{\ast}}.
\end{equation}
The first term in Eq.~(\ref{eq8}) describes elastic scattering, while $S_{vib}^{(1)}(q,\omega)$
describes inelastic scattering from the vibrational normal modes in the
one-mode approximation. Multimode scattering will arise from the higher
order terms in  Eq.~(\ref{eq4}), and are neglected in the present work. It is
understood that the three modes of uniform translation, for which $\omega_{\lambda} = 0$,
are omitted from all statistical mechanics equations.

We now allow for transits. When atom $K$ is involved in a transit,
both $\bm{R}_{K}$ and $\bm{u}_{K}$ change in a very short time, in such a way that 
$\bm{R}_{K} + \bm{u}_{K}$ 
remains continuous and differentiable in time. A detailed model of transits
in the atomic trajectory may be found in Chisolm et al. \cite{ChCWPRE63,ChWJPCM01}. Here we
seek a simpler approximation. If the time segments between transits involving
atom $K$ are denoted $\gamma_{K}=1,2,\dots$, then the position of atom $K$ at time $t$ is
$\bm{R}_{K}(\gamma_{K}(t)) + \bm{u}_{K}(\gamma_{K}(t),t)$. $F(q,t)$ for the liquid is then
written, from Eq.~(\ref{eq1}),
\begin{equation} \label{eq11}
F_{liq}(q,t) = \frac{1}{N}
 \left <\sum_{KL} e^{-i \bm{q}\cdot [\bm{R}_{K}(\gamma_{K}(t))-\bm{R}_{L}(t=0)]} 
\;e^{-i \bm{q} \cdot [\bm{u}_{K}(\gamma_{K}(t),t) - \bm{u}_{L}(t=0)]} \right >.
\end{equation}
Our numerical studies provide evidence, described below in connection with
Eq.~(\ref{eq14}), that transits can be approximately neglected in the displacements
$\bm{u}_{K}(\gamma_{K}(t),t)$. We therefore make this approximation, and separately
average the displacement terms in Eq.~(\ref{eq11}) over harmonic vibrations. The 
result is Eq.~(\ref{eq4}) with $\bm{R}_{K}$ replaced by $\bm{R}_{K}(\gamma_{K}(t))$,
\begin{equation}\label{eq12}
F_{liq}(q,t) \approx
\frac {1}{N} \Bigg < \sum_{KL}e^{-i \bm{q}\cdot [\bm{R}_{K}(\gamma_{K}(t))-\bm{R}_{L}(0)]}
\; e^{-W_{K}(\bm{q})}e^{-W_{L}(\bm{q})}
\;\left[ 1 + \left < \bm{q} \cdot \bm{u}_{K}(t)\;\bm{q} \cdot \bm{u}_{L}(0)\right >_{h}+ \dots\right]
                     \Bigg >_{\bm{q}^{\ast}} .
\end{equation}
Our next step is to modify this equation so as to model the presence of transits in 
$\bm{R}_{K}$.

There are two ways in which transits contribute to $F_{liq}(q,t)$. First,
transits introduce a fluctuating phase in the complex exponential in Eq.~(\ref{eq12}),
and this causes additional time decay through decorrelation along each
atomic trajectory. We model this with a relaxation function of the form
$e^{-\alpha t}$. Second, transits give rise to inelastic scattering, in addition to the
vibrational mode scattering already present in Eq.~(\ref{eq12}), and this increases
the total scattering cross section. We model this with a multiplicative
factor.

The leading term in Eq.~(\ref{eq12}) gives rise to the liquid Rayleigh peak, and
so is denoted $F_{R}(q,t)$. Without transits, $F_{R}(q,t)$ reduces to $F_{vib}(q,\infty)$,
Eq.~(\ref{eq6}), so we model $F_{R}(q,t)$ as 
\begin{equation} \label{eq13}
F_{R}(q,t) = C(q)\; F_{vib}(q,\infty)\; e^{-\alpha_{1}(q)\; t}.
\end{equation}
This function decays to zero with increasing time, in accord with the liquid
property $F_{liq}(q,t) \rightarrow 0$ as $t \rightarrow \infty$. 
The relaxation rate $\alpha_{1}(q)$ is expected to
be around the mean single-atom transit rate. $C(q)$ is positive, and greater than $1$
because of the inelastic scattering associated with transits (notice the total scattering
cross section is not affected by the factor $e^{-\alpha_{1} t}$).

The displacement-displacement correlation function in Eq.~(\ref{eq12}) gives rise to
the Brillouin peak, and so is denoted $F_{B}(q,t)$. Without transits $F_{B}(q,t)$
reduces to $F_{vib}(q,t)-F_{vib}(q,\infty)$, Eq.~(\ref{eq7}). Empirically, we have found that
this vibrational contribution alone gives an excellent account of the location
of the Brillouin peak, and the total cross section within it, as compared
with MD calculations and with experimental data for liquid sodium \cite{ARXIV05}. 
This suggests  keeping the vibrational contribution intact, as we did in
going to Eq.~(\ref{eq12}), and also suggests that we model $F_{B}(q,t)$ by

\begin{equation} \label{eq14}
F_{B}(q,t) = [F_{vib}(q,t) - F_{vib}(q,\infty)]\;e^{-\alpha_{2}(q) \;t}.
\end{equation}
Note $F_{vib}(q,t) - F_{vib}(q,\infty)$ decays to zero with time, because of the decay of
the vibrational correlation function in Eq.~(\ref{eq7}), and this decay gives the  
Brillouin peak its natural width \cite{ARXIV05}. The right side of Eq.~(\ref{eq14})
decays faster with time, hence broadens the Brillouin peak from its natural width,
but leaves its total cross section unchanged.

\begin{figure}[h]
\includegraphics[height=3.0in,width=3.0in]{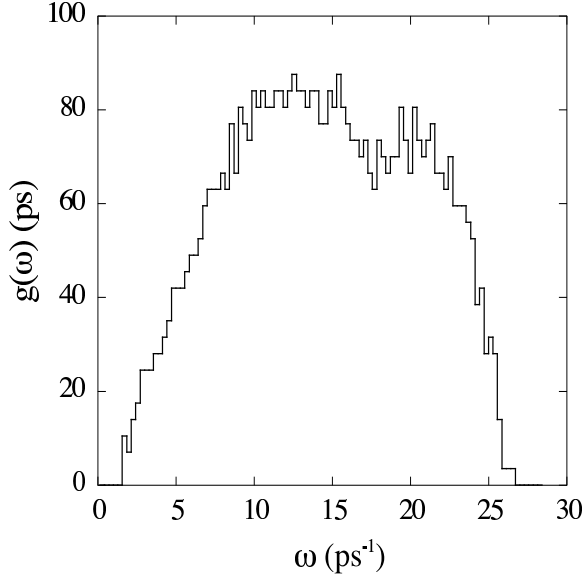}
\caption { Vibrational mode frequency distribution for a single random valley (1497 modes).} 
\end{figure}

From the above equations, our model for the dynamic structure factor is
\begin{equation} \label{eq15}
S_{liq}(q,\omega)=S_{R}(q,\omega)+S_{B}(q,\omega),
\end{equation}
\begin{equation}\label{eq16}
S_{R}(q,\omega) = \frac{C(q)\; \alpha_{1}\; F_{vib}(q,\infty)}{\pi(\omega^{2}+\alpha_{1}^{2})},
\end{equation}
\begin{equation}\label{eq17}
S_{B}(q,\omega) = \frac{3kT}{2M}\frac{1}{3N}\sum_{\lambda}f_{\lambda}(q)\frac{1}{\pi}
\left [ \frac{\alpha_{2}}{(\omega +\omega_{\lambda})^{2}+\alpha_{2}^{2}} +
 \frac{\alpha_{2}}{(\omega -\omega_{\lambda})^{2}+\alpha_{2}^{2}}\right ].
\end{equation}
The model has three adjustable parameters for each $q$, namely $C(q), \alpha_{1}(q)$, and
$\alpha_{2}(q)$.

We note the presence of a short-time error in $F_{liq}(q,t)$. The correct short-time
behavior is $F_{liq}(q,0) + b\;t^{2} + \dots$, with known coefficient $b$ (\cite{HMCDbook}, Eq.~(7.4.41)). 
The vibrational contribution, Eq.~(\ref{eq7}), has the correct
limiting behavior, but the model functions in Eq.~(\ref{eq13}) and (\ref{eq14}) are linear in
time, since $e^{-\alpha t} = 1 - \alpha t + \dots$. The linear term is important up to a time
$t_{C}$, which is very small, and beyond $t_{C}$ the time dependence of $F_{vib}(q,t)$
dominates. In our system we estimate the linear time dependence
contributes to $S_{liq}(q,\omega)$ only at frequencies above $50$ps$^{-1}$, which is above
the largest $\omega_{\lambda}$ present. For reference, the normal mode frequency
distribution $g(\omega)$ is shown in Fig.~1.

To complete this Section, let us estimate the average rate $\nu$ at which an individual atom is involved in 
a transit. From studies of the velocity autocorrelation function \cite{DWPRE58,ChCWPRE63}, our general estimate 
for monatomic liquids at melt is
$\nu \approx \left<\omega\right>/2\pi$, where $\left<\omega\right>$ is the rms vibrational mode frequency. For liquid
sodium at melt, this gives $\nu \approx 2.5$ ps$^{-1}$.

\section{Results and Discussion}

The system we study has $N=500$ atoms with an interatomic potential
representing metallic sodium at the density of the liquid at melt. The
potential gives an accurate account of the vibrational and thermodynamic
properties of crystal and liquid phases, and a good account of self diffusion
in the liquid (for summaries see \cite{DWbook2,ChWJPCM01}). 
Here we use the sodium potential to see how well our transit model
can be made to fit MD results for $S(q,\omega)$. We study $q$-values in the range
from $0.12$~a$_{0}^{-1}$, the smallest allowed $q$ for our system, up to $0.81$~a$_{0}^{-1}$, beyond
which the Brillouin peak is poorly discernable. In comparison, the
first peak in $S(q)$ is at $q_{m}=1.05$~a$_{0}^{-1}$. The model is evaluated from
Eqs.~(\ref{eq15}-\ref{eq17}) for a single random valley, and the results show scatter due to
the small system size. To reduce this scatter we used a graphically smoothed
curve of $F_{vib}(q,\infty)$ in Eq.~(\ref{eq16}) for $S_{R}(q,\omega)$, where the smoothed data are listed
in Table~I. In comparison, the MD results show little finite-$N$ scatter, since
the MD system visits a very large number of random valleys during the decay time 
of $F(q,t)$.

\begin{table}
\caption{\label{table1} Quantities associated with our liquid $S(q,\omega)$ model.}
\begin{ruledtabular}
\begin{tabular}{crccc}
$\bm{q}$ & $q$~(a$_{0}^{-1}$) & $F_{vib}(q,\infty)$ & $F_{vib}^{(1)}(q,0)/F_{vib}(q,0)$ & $C(q)$\\
\hline
(0,0,1)          & 0.12129  & 0.0043  & 1.00 & 2.0\\
(1,1,1)          & 0.21009  & 0.0033  & 0.98 & 2.3\\
(1,1,2)          & 0.29711  & 0.0026  & 0.94 & 2.9\\
(1,1,3)          & 0.40229  & 0.0021  & 0.90 & 4.0\\
(0,1,4),(2,2,3)  & 0.50011  & 0.0021  & 0.84 & 5.1\\
(0,3,5),(3,3,4)  & 0.70726  & 0.0044  & 0.72 & 8.3\\
(0,3,6),(2,4,5)  & 0.81367  & 0.0110  & 0.70 & 9.4\\
\end {tabular}
\end{ruledtabular}
\end{table}

\begin{figure}[h]
\includegraphics[height=3.0in,width=3.0in]{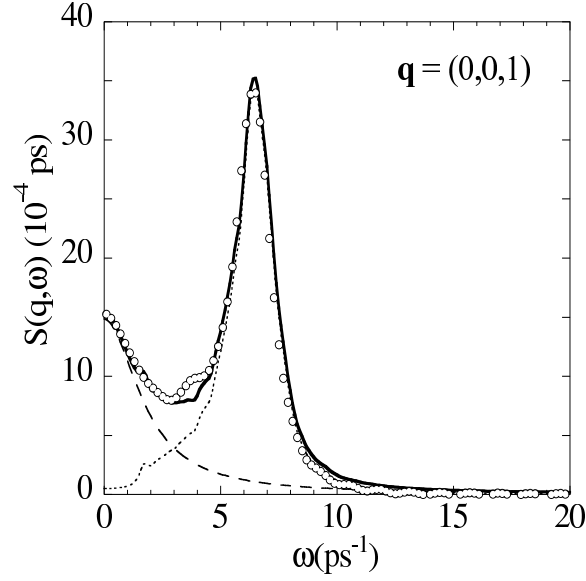}
\caption {\label{S001} $S(q,\omega)$ for $q=0.12129$~a$_{0}^{-1}$, from the model (solid line)
 and from MD (circles). The Rayleigh (broken line) and the  Brillouin  (dotted line) contributions 
 to the model are also shown separately.}
\end{figure}

The individual functions $S_{R}(q,\omega)$, $S_{B}(q,\omega)$, their sum $S_{liq}(q,\omega)$, and
$S_{MD}(q,\omega)$ are shown in Fig.~2 for a representative $q$. The slight decrease in
$S_{B}(q,\omega)$ at small $q$ appears because our system has no vibrational modes
with frequencies below 1.7~ps$^{-1}$ (see Fig.~1). In the fitting process, we adjusted $C$ and $\alpha_{1}$
to get fits of the intercept at $\omega=0$, and of the slope in the steeply decreasing
range at $\omega \approx$ 1--3~ps$^{-1}$, and we adjusted $\alpha_{2}$ to get an overall fit to the Brillouin peak.

The fitted $S_{liq}(q,\omega)$ and $S_{MD}(q,\omega)$ are shown for the remaining $q$-values in
Figs.~3-8. For each $q$, the shapes of the components $S_{R}(q,\omega)$  and $S_{B}(q,\omega)$ 
are qualitatively the same as those shown in Fig.~2. The overall fits of
our model to MD data are very good for $q$ from 0.12 to 0.50~a$_{0}^{-1}$, from Figs.~2-6. 
Except for the small undershoot of theory at the bottom of the dip
between Rayleigh and Brillouin peaks, and the overshoot of theory in the
 high frequency tail, the model discrepancies can be attributed to scatter
 due to evaluation for only one random valley. But at $q$ around 0.71 and 0.81~a$_{0}^{-1}$,
 Figs.~7 and 8, the model cannot be made to fit the Brillouin peak quite so well as in the 
 preceeding five figures. The figures suggest that our overall physical description
 is still correct at the two highest $q$,
 but a small correction needs to be addressed there.

\begin{figure}[h]
\includegraphics[height=3.0in,width=3.0in]{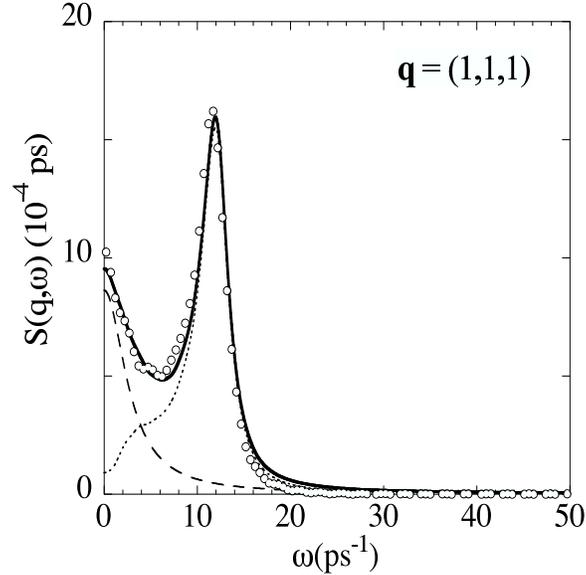}
\caption {\label{S111}Same as Fig.~2 but for $q=0.21009$~a$_{0}^{-1}$.}
\end{figure}

\begin{figure}[h]
\includegraphics[height=3.0in,width=3.0in]{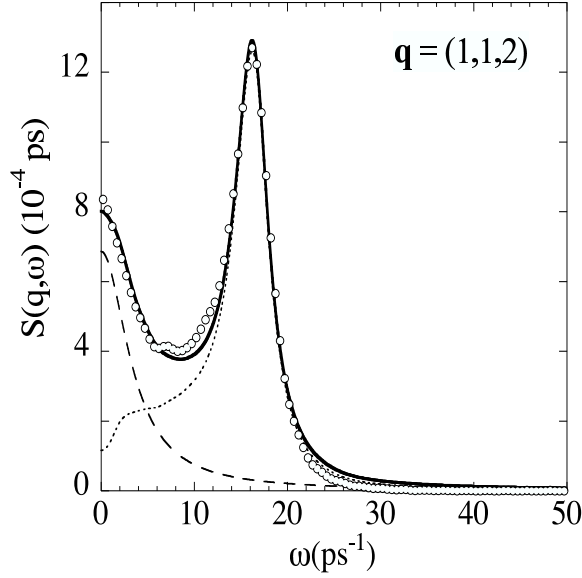}
\caption {\label{S112}Same as Fig.~2 but for $q=0.29711$~a$_{0}^{-1}$.}
\end{figure}

\begin{figure}[h]
\includegraphics[height=3.0in,width=3.0in]{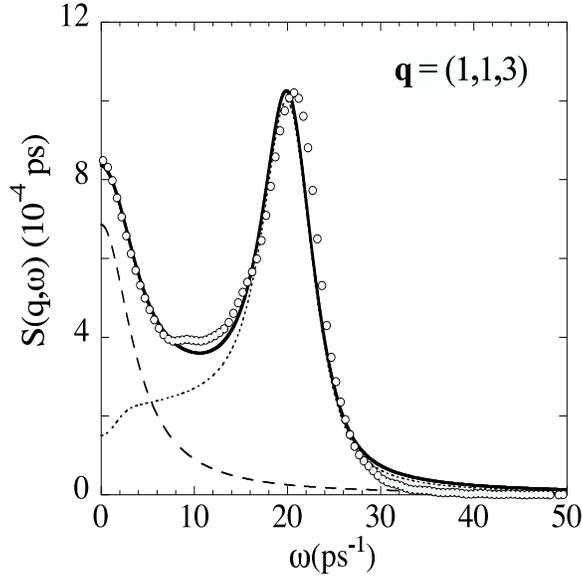}
\caption {\label{S113} Same as Fig.~2 but for $q=0.40029$~a$_{0}^{-1}$.}
\end{figure}

\begin{figure}[h]
\includegraphics[height=3.0in,width=3.0in]{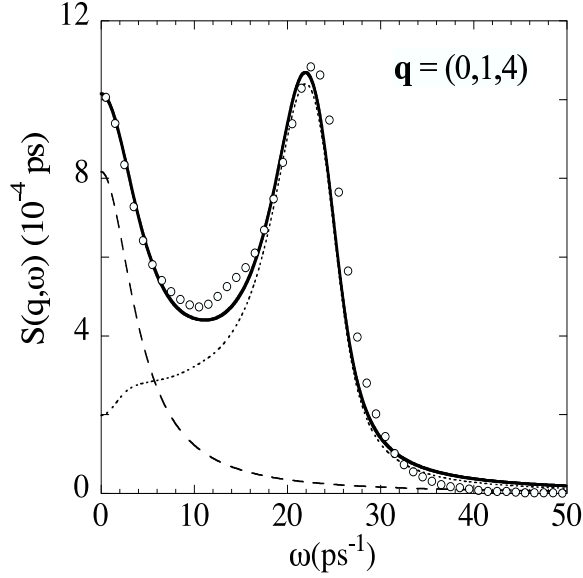}
\caption {\label{S014} Same as Fig.~2 but for  $q=0.50011$~a$_{0}^{-1}$.}
\end{figure}

\begin{figure}[h]
\includegraphics[height=3.0in,width=3.0in]{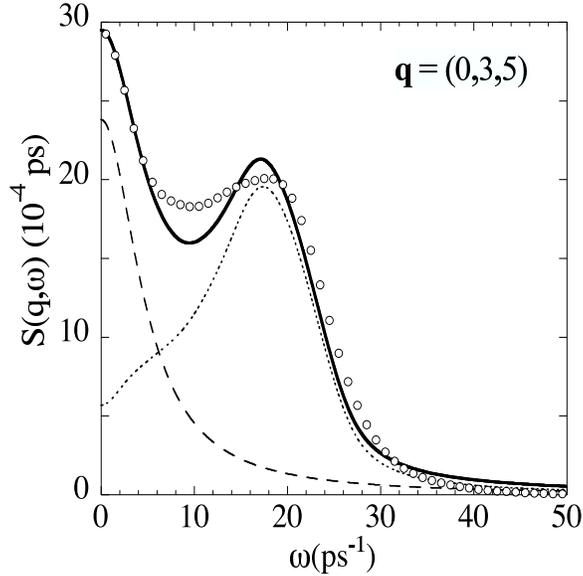}
\caption {\label{S035} Same as Fig.~2 but for $q=0.70726$~a$_{0}^{-1}$.}
\end{figure}

\begin{figure}[h]
\includegraphics[height=3.0in,width=3.0in]{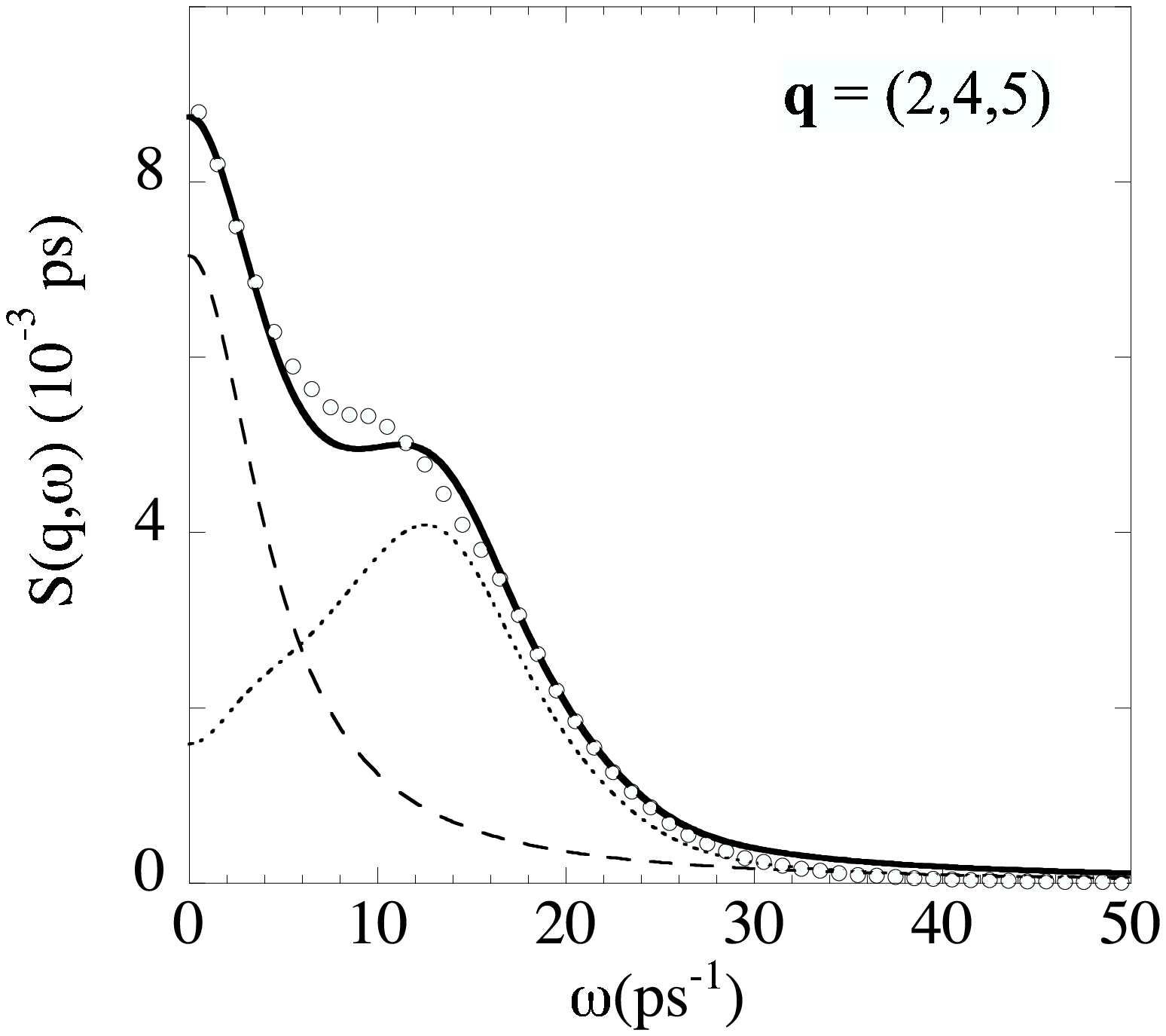}
\caption {\label{S245} Same as Fig.~2 but for $q=0.81367$~a$_{0}^{-1}$.}
\end{figure}

 Experience with inelastic neutron scattering in crystals suggests that
 multimode scattering should become significant at temperatures close to melting and $q$ beyond the
 first Brillouin zone boundary. This includes our system at $q\gtrsim 0.70$~a$_{0}^{-1}$.
 We note that $F_{vib}(q,\infty)$, Eq.~(\ref{eq6}), is not affected by the one-mode
 approximation. Since $F_{vib}(q,0)$ is the maximum magnitude of $F_{vib}(q,t)$,
 an estimate of the accuracy of the one-mode approximation is provided by 
 the ratio $F_{vib}^{(1)}(q,0)/F_{vib}(q,0)$, where the numerator is the one-mode approximation
 and the denominator is exact. The ratio is listed for each $q$ in Table~I. We
 interpret the results as follows. Multimode scattering, not included  in
 Eq.~(\ref{eq17}) for $S_{B}(q,\omega)$, is present in the MD data, is small for 
$q\lesssim  0.50$~a$_{0}^{-1}$, but is responsible for the discrepancy
between model and MD around the Brillouin peak in Figs.~7 and 8.

Let us now consider the magnitude of the fitted parameters. The Rayleigh
peak strength $C(q)$, listed in Table I, is greater than 1 and increases steadily
as $q$ increases. This implies a considerable cross section for inelastic transit
scattering. The Rayleigh peak relaxation rate $\alpha_{1}(q)$ is graphed in Fig.~9. 
From Eq.~(\ref{eq16}), $\alpha_{1}(q)$ is the Rayleigh peak half width at half max. $\alpha_{1}(q)$
extrapolates toward zero as $q \rightarrow 0$, while  $\alpha_{1}(q)$ is roughly constant at
$q\gtrsim  0.5$~a$_{0}^{-1}$. The Brillouin peak relaxation rate $\alpha_{2}(q)$ is also graphed in
Fig.~9, and appears to go to zero at $q$ around 0.1~a$_{0}^{-1}$. Notice $\alpha_{2}(q)$ does
not measure the width of the Brillouin peak, but measures its width
beyond the natural width. Hence the $\alpha_{2}(q)$ graph suggests that the
liquid Brillouin peak has its natural (vibrational only) width at small
$q$, at $q\lesssim 0.1$~a$_{0}^{-1}$ in the present work. Except where the relaxation
rates approach zero at small $q$, $\alpha_{1}$ and $\alpha_{2}$ are in the range 
1--5~ps$^{-1}$,  in qualitative agreement with the mean transit rate $\nu \approx 2.5$~ps$^{-1}$.

\begin{figure}[h]
\includegraphics[height=3.0in,width=3.0in]{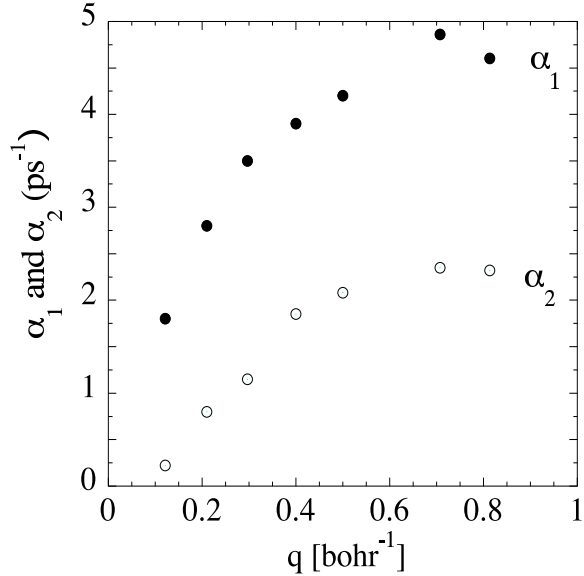}
\caption {\label{avsq} Relaxation rates from our liquid $S(q,\omega)$ model: $\alpha_{1}$  (filled circles) for the Rayleigh peak
and  $\alpha_{2}$ (empty circles) for the Brillouin peak.}
\end{figure}

\section{Comparison with Mode Coupling Theory}

The theory most successful to date in accounting for time correlation functions 
is mode coupling theory \cite{GoLu75}. Detailed summaries of mode coupling theories of liquid
dynamics are given by Boon and Yip \cite{BYbook}, Hansen and McDonald \cite{HMCDbook}, and
Balucani and Zoppi \cite{BZbook}. Mode coupling theory has been applied to the glass
transition \cite{BGS84,Leut84,Gotze99}, and has been shown capable of rationalizing the density
correlation functions at temperatures in the vicinity of the glass transition 
\cite{Ka94,KA95a,KA95b}. The application for which we shall compare mode coupling and V-T theories
is dynamic response of monatomic liquids at temperatures near and above melting.

Mode coupling theory works with the generalized Langevin equation for $F(q,t)$,
and expresses the memory function in terms of processes through which
density fluctuations decay \cite{BYbook, HMCDbook, BZbook}.  In the viscoelastic approximation, the
memory function decays with a $q$-dependent relaxation time \cite{BYbook, HMCDbook, BZbook}. This
approximation provides a good fit to the combined experimental data \cite{CRPRL74} and
MD data \cite{ARPRL74,ARPRA74} for the Brillouin peak dispersion curve in liquid Rb \cite{CLRPP38} (see also Fig.~9.2
of \cite{HMCDbook}). Going beyond the viscoelastic approximation,
Bosse et al \cite{BGLPRA17a, BGLPRA17b} constructed a self-consistent theory for the
longitudinal and transverse current fluctuation spectra, each expressed in
terms of relaxation kernels approximated by decay integrals which couple
the longitudinal and transverse excitations. This theory is in good overall
agreement with extensive neutron scattering data and MD calculations
for Ar near its triple point \cite{BGLPRA17b}. The theory was developed further by
Sj\"ogren \cite{Sjog80, Sjog80b}, who separated the memory function into a binary
collision part, approximated with a Gaussian ansatz, and a more collective
tail represented by a mode coupling term. For liquid Rb, this theory
gives an ``almost quantitative'' agreement with results from neutron scattering experiments
\cite{CRPRL74} and MD calculations \cite{ARPRL74,ARPRA74}. More recently, inelastic x-ray scattering 
measurements have been done for the
light alkali metals Li \cite{R&c002} and Na \cite{SBRSb,Naexp}. 
These data have been analyzed by mode coupling theory, and the resulting 
fits to $S(q,\omega)$ are excellent, both for the experimental data and for MD
calculations \cite{SBRSb,SRSVPRE02,SRSVPM02,SBRS,SBRSa}.

A detailed comparison of the present study with the analysis of Scopigno et al.
\cite{SBRSb} is of interest. They analyzed experimental data for liquid sodium at 
390~K, for $q$ in the range 0.08--0.77~a$_{0}^{-1}$. Their memory function has three relaxation
terms, one for coupling between thermal and density degrees of freedom, with no
adjustable parameters, and two for true viscous processes, each having
adjustable weight and relaxation time. One parameter is fixed by the total
weight, so Scopigno et al. have effectively three $q$-dependent parameters to fit the shape
of $S(q,\omega)$. In the present work, we calibrate V-T theory by comparison with MD
data for liquid sodium at 395~K, for $q$ in the range  0.12--0.81~a$_{0}^{-1}$. Our 
Rayleigh peak contribution has one weight parameter and one relaxation rate,
and our Brillouin peak contribution has one relaxation rate, making three adjustable 
$q$-dependent parameters. Our Brillouin peak weight parameter was fixed at one because 
we had already confirmed that the vibrational contribution
alone has the correct weight. We find that our $\alpha_{1}$ is nearly the same
as Scopigno et al.'s inverse relaxation time $\tau_{\alpha}^{-1}$  over the entire $q$ range: the
relative difference in magnitude averages 25$\%$. Though we understand
the reason, that in each case the parameter is determined by the width of
the Rayleigh peak, the level of agreement is remarkable nevertheless. On
 the other hand, our $\alpha_{2}$ is \textit{much} smaller than their $\tau_{\mu}^{-1}$,
 as the ratio $\alpha_{2}/\tau_{\mu}^{-1}$ varies from 0.01 at small $q$ to 0.07 at large $q$. 
 Again the reason is clear: while $\tau_{\mu}^{-1}$ is determined by the Brillouin peak width, $\alpha_{2}$ is
 determined by only the width beyond the natural width. The result illustrates
 the important point of comparison between V-T and mode coupling theories: 
 the two methods are based on different decompositions of the physical processes
 involved. While mode coupling theory analyzes $F(q,t)$ in terms of processes
 by which density fluctuations decay, V-T theory analyzes $F(q,t)$ in terms of
 the two contributions to the total liquid motion, vibrations and transits.

 \section{Conclusions}
 
 By resolving the complete atomic motion into its constituents, vibrations and 
 transits, V-T theory offers a unified  theoretical formulation of equilibrium
 statistical mechanics averages, both of thermodynamic variables and of time
 correlation functions. The vibrations alone provide a tractable, accurate,
 parameter free formulation of thermodynamic properties of monatomic liquids \cite{DWPRE56}.
 The role of transits is merely to allow the liquid to visit the vast array 
 of random valleys, and hence achieve the full liquid entropy \cite{DWPRE56,ChWJPCM01,DWbook2}. The same
 vibrations provide a tractable, parameter free contribution to time correlation
 functions, while the same transits, which are part of the equilibrium
 fluctuations, complete the formulation of time correlation functions.
 
 Our study of the dynamic structure factor exemplifies this unification. The
 vibrational contribution provides the nontrivial results in Eqs.~(\ref{eq8}-\ref{eq10}) for
 $S_{vib}(q,\omega)$. Pure elastic scattering is given by $F_{vib}(q,\infty)\;\delta(\omega)$, where 
$F_{vib}(q,\infty)$ is the positive long-time limit of $F_{vib}(q,t)$.
 Inelastic scattering in the one-mode approximation is given by $S_{vib}^{(1)}(q,\omega)$,
a sum over independent normal-mode cross sections. $S_{vib}^{(1)}(q,\omega)$ provides the
natural width of the Brillouin peak \cite{ARXIV05}. For liquid sodium at melt,
$S_{vib}^{(1)}(q,\omega)$ gives a highly accurate account of the location of the Brillouin peak
\cite{ARXIV05}, and as we have seen in the present study, gives an accurate account
of the Brillouin peak area as well.

As shown in Sec.~II, transits contribute to $F_{liq}(q,t)$ in two ways. Transit-induced
jumps in the atomic equilibrium positions and displacements cause decorrelation
among the terms in Eq.~(\ref{eq11}), hence transits enhance the decay of time
correlations. Transits also provide an additional source of inelastic scattering,
hence increase the inelastic cross section. These effects are modeled by the 
strength parameter $C(q)$ and the relaxation function $e^{-\alpha_{1}(q)\;t}$ in Eq.~(\ref{eq13}) for
$F_{R}(q,t)$, and by the relaxation function $e^{-\alpha_{2}(q)\;t}$ in Eq.~(\ref{eq14}) for
$F_{B}(q,t)$. The
model so constructed is a generalization of Zwanzig's model for the velocity 
autocorrelation function \cite{ZJCP79}. The model expressions for $S_{liq}(q,\omega)$ are given in
 Eqs.~(\ref{eq15}-\ref{eq17}).
 
 As shown in Figs.~2-8, the model can be made to fit MD calculations of $S(q,\omega)$
 extremely well, almost within computational errors, except for the two largest $q$
 values. The small inadequacy of the model in the vicinity of the Brillouin peak
  in Figs.~7 and 8 is apparently due to multimode scattering, present in the MD
  calculation but not in the vibrational theory evaluated here. Properties of
  the fitting parameters are as follows. In the Rayleigh peak contribution, $C(q)$ is
  greater than one (Table~I), indicating the presence of inelastic transit scattering
  in the MD calculations. The relaxation rate $\alpha_{1}(q)$ is close to the mean transit
  rate of 2.5~ps$^{-1}$ as expected (Fig.~9), but the figure suggests that $\alpha_{1}(q)$
  approaches zero as $q \rightarrow 0$. The Brillouin peak relaxation rate $\alpha_{2}(q)$ appears to
  vanish at small $q$ (Fig.~9), which would mean that the Brillouin peak width
  in the liquid is the natural width at small $q$.

   Finally, while mode coupling theory and V-T theory each provide a 
   physically based model capable of accurately fitting $S(q,\omega)$ for the liquid 
   at melt, the two models work with entirely different projections of the 
   underlying liquid motion. V-T theory is the more universal, in that it applies the same motional 
   constituents to all statistical averages, equilibrium and nonequilibrium alike.

\acknowledgments{Eric Chisolm is gratefully acknowledged for collaboration and 
for critically reading the manuscript.}

\bibliography{Modelref} 

\end{document}